\documentclass[twocolumn]{aastex61}

\usepackage{natbib}

\shorttitle{GeV emission from HESS J1534-571}
\shortauthors{Araya, M.}

\begin{document}

\title{Detection of GeV gamma-rays from HESS J1534-571 and multiwavelength implications for the origin of the non-thermal emission}

\correspondingauthor{Miguel Araya}
\email{miguel.araya@ucr.ac.cr}

\author{Miguel Araya}
\affiliation{Escuela de F\'isica \& Centro de Investigaciones Espaciales (CINESPA),\\
Universidad de Costa Rica, San Jos\'e 2060,\\
Costa Rica}

\begin{abstract}
HESS J1534-571 is a very high energy gamma-ray source that was discovered by the H.E.S.S. observatory and reported as one of several new sources with a shell-like morphology at TeV energies, matching in size and location with the supernova remnant G323.7-1.0 discovered in radio observations by the Molonglo Galactic Plane Survey. Many known TeV shells also show X-ray emission, however, no X-ray counterpart has been seen for HESS J1534-571. The detection of a new GeV source using data from the \emph{Fermi} satellite that is compatible in extension with the radio supernova remnant and shows a very hard power-law spectrum ($\frac{dN}{dE} \propto E^{-1.35}$) is presented here, together with the first broadband modeling of the nonthermal emission from this source. It is shown that leptonic emission is compatible with the known multiwavelength data and a corresponding set of physical source parameters is given. The required total energy budget in leptons is reasonable, $\sim 1.5\times10^{48}$ erg for a distance to the object of $5$ kpc. The new GeV observations imply that a hadronic scenario, on the other hand, requires a cosmic ray spectrum that deviates considerably from theoretical expectations of particle acceleration.
\end{abstract}


\section{Introduction} \label{sec:intro}
Supernova remnants (SNRs) have been considered the main sources of Galactic cosmic rays for a long time from energetic considerations. Charged particles are thought to gain energy by the mechanism of diffusive shock acceleration \citep[e.g.,][]{1977DoSSR.234.1306K,1978MNRAS.182..147B} in the blast waves of supernova explosions. The process of cosmic ray scattering is thought to be enhanced by the presence of magnetic field fluctuations \citep[e.g.,][]{1978MNRAS.182..147B,2004MNRAS.353..550B} that could allow some particles to reach PeV energies and still be able to remain confined within the sources. Observations of several SNRs at MeV-GeV energies with the \emph{Fermi} Large Area Telescope \citep[LAT,][]{2009ApJ...697.1071A} have revealed the characteristic gamma-ray spectrum expected from the decay of neutral pions produced in hadronic interactions in several sources located near or interacting with high-density material \citep{2013Sci...339..807A,2016ApJ...816..100J}. These sources are middle-aged ($\sim 10^4$ yr) and show a break in the spectrum at GeV energies.

Other types of SNRs that produce gamma-rays are TeV shells such as RX J1713.7-3946, RCW 86 and RX J0852.0-4622, which show hard power-law GeV spectra of the form $E^{-\Gamma}$, with $\Gamma \sim 1.5$ and $1.4$ in the first two cases. Such a spectral shape can naturally be explained by interactions of high-energy leptons with ambient photon fields through the process of inverse Compton scattering (IC) \citep[e.g.,][]{2015A&A...580A..74A}. For other types of SNRs, mixed contributions from hadronic emission and leptonic processes (including perhaps considerable non-thermal bremsstrahlung) are possible \citep[e.g.,][]{2010ApJ...720...20A}.

For the synchrotron-dominated TeV shell RX J1713.7-3946, which shows very low thermal X-ray emission and thus low average gas densities, it has been proposed that hadronic processes could account for the hard gamma-ray spectrum when very energetic cosmic rays penetrate dense and cold clumps of matter located within the shell \citep[see, e.g.,][]{1996A&A...309..917A,2012ApJ...746...82F,2014MNRAS.445L..70G,2017arXiv170200664T}. In order to identify the nature of the gamma-rays from SNRs constraining the values of ambient parameters and properties is crucial, and this can be done through multiwavelength observations.

HESS J1534-571 was discovered at TeV energies by the High Energy Stereoscopic System (H.E.S.S.), an imaging atmospheric Cherenkov telescope \citep{2015arXiv150903872P}. It was classified as a TeV-SNR of the shell type from its morphology. The associated SNR candidate, G323.7-1.0, was also recently discovered in radio observations of the Molonglo Galactic Plane Survey, which revealed an extremely faint $\sim 0.9\degr \times 0.6\degr$ oval shell \citep{2014PASA...31...42G} that matches the TeV emission. \cite{2015arXiv150903872P} note that no X-ray emission in the region of HESS J1534-571 has been detected from the \emph{ROSAT} survey data, and this could be explained from Galactic absorption. However they also note that no X-ray emission has been seen in several \emph{Suzaku} observations that covered parts of the shell. The absence of X-ray synchrotron emission would make HESS J1534-571 atypical as other TeV shells such as RX J1713.7-3946, SN 1006, RCW 86, RX J0852.0-4622 and HESS J1731-347, show non-thermal X-ray emission \citep[e.g.,][]{2016PASJ...68S..10T,2016ApJ...819...98A,2012MNRAS.425.2810A,2016arXiv161200258C,2015ApJ...799..175S}. There are, however, TeV shell candidates with no known X-ray counterpart such as HESS J1614-518 \citep{2011PASJ...63S.879S} and HESS J1912+101 \citep{2015arXiv150903872P}.

In this work, the detection of an extended GeV source with the LAT at the location of the SNR G323.7-1.0 is shown. No previous source has been reported at this location in the LAT Third Source Catalog, the latest catalogs of hard LAT sources nor the First Fermi LAT Supernova Remnant Catalog \citep{2015ApJS..218...23A,0067-0049-222-1-5,2016ApJS..224....8A,2017arXiv170200664T,2017arXiv170200476T}. The source extension matches the radio shell and it shows a hard GeV spectrum. The combined gamma-ray observations from the \emph{Fermi} LAT satellite and preliminary H.E.S.S. fluxes can be explained within the framework of a simple one-zone leptonic model and a set of possible source parameter values is derived. Finally, a hadronic scenario for the GeV-TeV emission is also studied, although this scenario is less likely since the required particle spectrum deviates considerably from standard predictions of cosmic ray acceleration.

\section{\emph{Fermi} LAT data} \label{sec:LAT}
The \emph{Fermi} LAT is a gamma-ray telescope that detects photons in the energy range between 20 MeV and $> 500$ MeV. It has a converter/tracker for direction measurement, a calorimeter for energy measurement and an anti-coincidence detector for charged background identification. The analysis presented here used data collected from the beginning of the mission (August 2008) to January 2017 and were analyzed with the recently improved effective area, point-spread function and energy resolution (Pass 8). Version v10r0p5 of the ScienceTools and instrument response functions with the event class SOURCE were used. Events were selected having a maximum zenith angle of 90$\degr$ and when the data quality was good. Data taken prior to 2008 October 27 have been excluded as they were affected by a higher level of background contamination, which is particularly important at the highest energies.

Events with energies from 5 GeV to 500 GeV were chosen within a 20$\degr \times 20 \degr$ region of interest around the position RA=233.5$\degr$, Dec=-57$\degr$ (J2000). The LAT tools fit a model to the data including the residual charged particles and diffuse gamma-rays. A maximum likelihood technique \citep{1996ApJ...461..396M} was used to obtain the relevant source parameters and significance. The Galactic diffuse model used is given by the file gll\_iem\_v06.fits and the residual background and extragalactic emission were modeled with an isotropic component given by the file iso\_P8R2\_SOURCE\_V6\_v06.txt, which are provided by the LAT team. The data are binned spatially in counts maps using a scale of 0.05$\degr$ per pixel. For exposure calculation, ten logarithmically spaced bins per decade in energy were used. The sources included in the LAT Third Source Catalog \citep[3FGL,][]{2015ApJS..218...23A} within the region of interest are considered in the fit, but only the normalizations of the sources located within 5$\degr$ around the center of the region were refitted with the increased data set, keeping their spectral shape fixed to the values reported in the catalog. The sources which are not significantly detected above 5 GeV were removed from the model. The detection significance was estimated as the square root of the test statistic (TS), defined as $-2\times$log$(L_0/L)$, with $L_0$ and $L$ the likelihood values without and with the corresponding source, respectively.

After subtracting the best-fit model to the data, residuals are seen at the location of the radio shell of G323.7-1.0. Additional residuals of extended emission, unrelated to this SNR, are seen around the positions RA=$238.46\degr$, Dec=$-53.4\degr$ and RA=$238.2\degr$, Dec=$-56.2\degr$ (J2000). This emission was accounted for with uniform disc templates of radii 0.5$\degr$ and 0.2$\degr$, respectively. The discs correspond to the sources 3FHL J1553.8-5325e and 3FHL J1552.7-5611e from the recently released Third Catalog of Hard \emph{Fermi}-LAT Sources \citep{2017arXiv170200664T}. The residuals map in the region of the SNR is shown in Fig. \ref{residuals}. The apparent residual emission extending to higher Galactic latitudes away from the SNR is not significant as demonstrated by fitting a point source at different locations near the region which results in a maximum TS of $\sim 2$.

\subsection{Source morphology}
Two spatial hypothesis to account for the emission were used for comparison, a uniform disc and ellipse. A spectral assumption of a simple power law ($\frac{dN}{dE} \propto E^{-\Gamma}$) for the emission was used for both spatial templates. For the uniform disc, a systematic search for the best fit template was carried out by moving the disc center along a $1\degr \times 1\degr$ grid centered near the geometrical center of the radio SNR, in steps of 0.1$\degr$, and changing its radius by 0.1$\degr$ in each step, from 0.1$\degr$ to 0.6$\degr$, and finding the maximum likelihood each time with the normalization and spectral index free in the fit. This results in five degrees of freedom in the fit. The uniform ellipse had a fixed shape tracing the radio border of the SNR, and thus the likelihood model with this morphology only had two degrees of freedom with respect to the null hypothesis, the normalization and spectral index.

\begin{figure}[ht!]
\includegraphics[width=9.5cm,height=6.3cm]{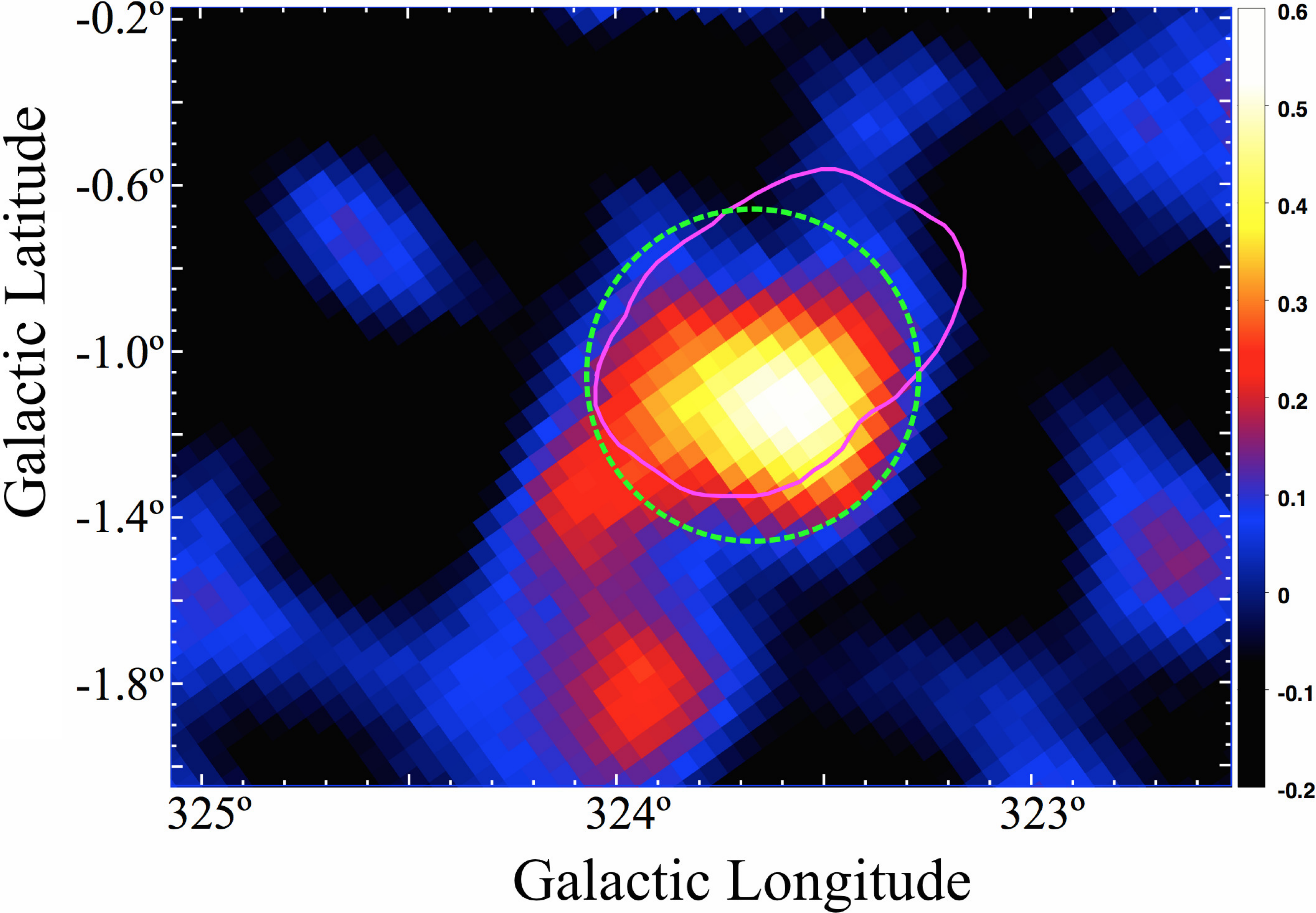}
\caption{Map of residual emission above 5 GeV in the region of the SNR G323.7-1.0 obtained after subtracting the best-fit model from the data. The map has units of counts per pixel and has been smoothed by a Gaussian with $\sigma= 0.15\degr$. The dashed circle represents the best-fit disc template found in Section \ref{sec:LAT} and the solid line shows the approximate radio border of the SNR, taken from \cite{2014PASA...31...42G}. No other sources from the LAT 3FGL catalog exist in this region.\label{residuals}}
\end{figure}

The maximum TS value found for the disc hypothesis was 57.4 with a corresponding disc radius of $0.4^{+0.2}_{-0.1}\degr$. The center of the best-fit template is located at the coordinates RA=$233.5\degr$, Dec=$-57.2\degr$. The disc border is also shown in Fig. \ref{residuals}. The TS value found for the ellipse was 50. For the rest of the analysis, the disc template was adopted in this work and systematic uncertainties in the spectral parameters were estimated to account for the choice of morphology.

\subsection{Spectral shape}
When fitting the spectrum to a power-law with an exponential cutoff the resulting spectral index and the TS value are the same as that from a simple power-law fit but the resulting cutoff energy is not constrained, as the fit gives $1.0 \pm 2.0$ TeV. Therefore the simple power-law shape is used as the best-fit spectral shape in the 5-500 GeV energy range.

The resulting spectral index and flux above 5 GeV obtained were then $1.34\pm 0.17_{\mbox{\tiny stat}} \pm 0.1_{\mbox{\tiny sys}}$ and $(2.2 \pm 0.6_{\mbox{\tiny stat}} \pm 0.5_{\mbox{\tiny sys}})\times 10^{-10}$ photons cm$^{-2}$ s$^{-1}$. The systematic errors are estimated by combining the individual systematic errors resulting from uncertainty in the Galactic diffuse emission model and the source morphology. For the first one the normalization of the diffuse component was changed by $\pm 6\%$ to estimate new spectral parameters of the source as typically done for analysis of LAT data \citep[e.g.,][]{2009ApJ...706L...1A}.

\begin{figure}[ht!]
\includegraphics[width=10cm,height=6cm]{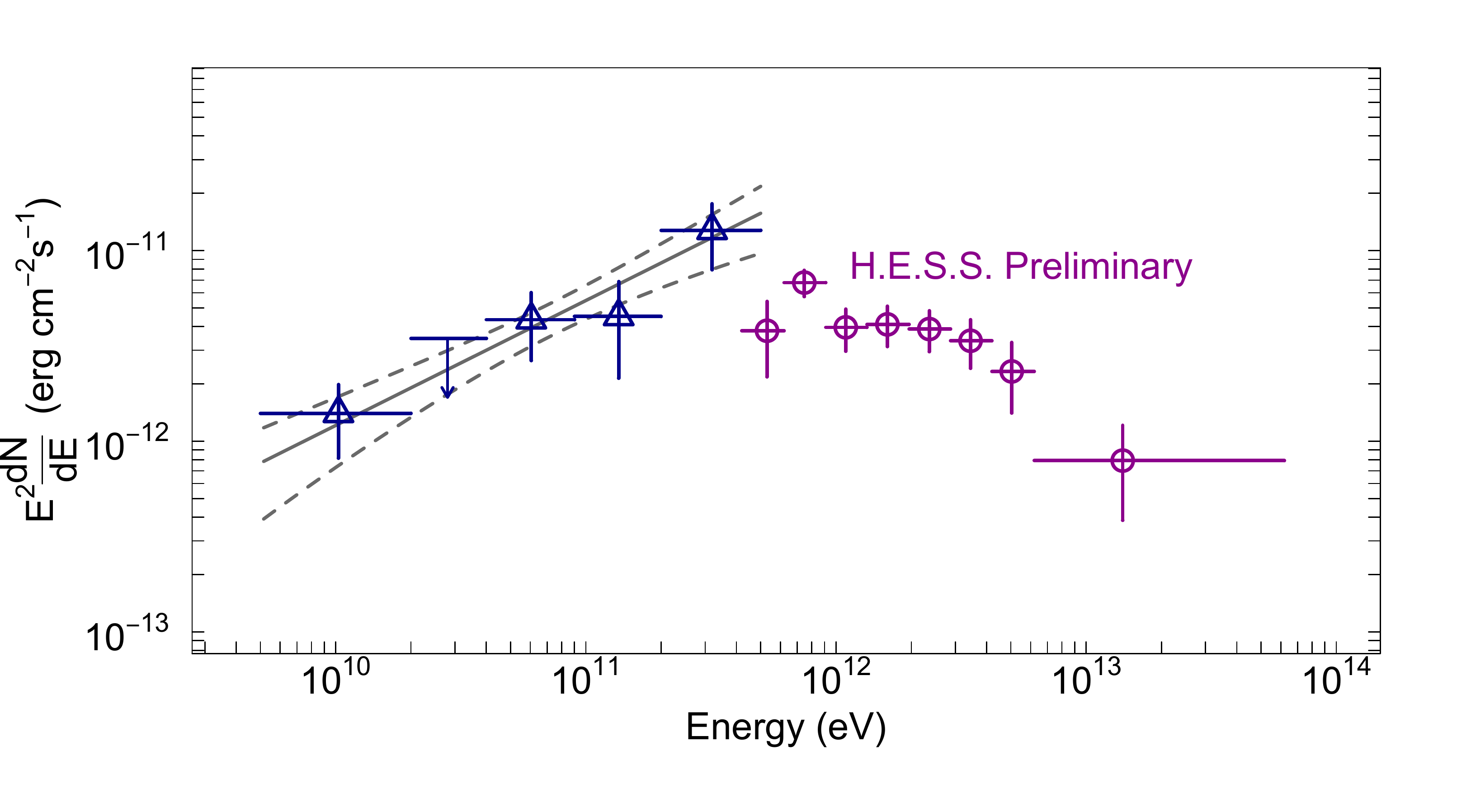}
\caption{Gamma-ray SED of HESS J1534-571. The solid line shows the best-fit spectrum in the 5--500 GeV energy range and the corresponding 1$\sigma$ statistical uncertainty is represented by the dashed lines. The LAT (triangles) and preliminary H.E.S.S. data points (circles) are also shown.\label{sed}}
\end{figure}

In order to probe the spectral behaviour in more detail the energy range was divided in five intervals and a likelihood fit was performed in each by keeping the values of the spectral parameters for all sources fixed to the ones obtained in the entire energy range, except for the normalization of the template representing HESS J1534-571. A value for the spectral energy distribution (SED) was estimated in each interval and a 95\%-confidence level upper limit was calculated for one of the intervals where the source TS is less than 7. The individual points are consistent with the fit for the whole data set, which supports the simple power-law spectral assumption. The data points and best-fit spectrum are shown in Fig. \ref{sed} together with the preliminary H.E.S.S. spectral points taken from \cite{2016arXiv161200261G}.

Finally, it is worth addressing the absence of a LAT source at the location of G323.7-1.0 in the recently released catalog of extended sources by \cite{2017arXiv170200476T}. Using the same data set as in their paper and the same cuts, a TS map was computed above 10 GeV in the region for a point source hypothesis. The map showed a maximum value of 20 located inside the SNR shell, which passes the threshold of 16 used in the paper to search for extended emission with a uniform disc hypothesis. Adding a disc at this location only increases the TS by 10, which is below the required value of 16 in their paper to claim a new extended source. The additional $\sim$40\% integration time compared to the analysis by \cite{2017arXiv170200476T} and the lower energy threshold used here can easily explain the new detection.

\section{Discussion} \label{sec:discussion}
Extended GeV emission from a source compatible with the size of the SNR G323.7-1.0 was found with a significance of 7.5$\sigma$ above 5 GeV in \emph{Fermi} LAT data. This source is likely the counterpart of the TeV shell HESS J1534-571. Just as is seen for other TeV shells, the GeV spectrum of this source is hard.

Using a recent parametrization of gamma-ray production cross sections for proton-proton interactions \citep{2014PhRvD..90l3014K}, the GeV-TeV data can be explained in a hadronic scenario with a particle distribution that is a power law with an exponential cutoff of the form $\propto \epsilon_p^{-s}\cdot e^{-\epsilon_p/ \epsilon_c}$, where $\epsilon_p$ is the hadron energy and the required spectral index and cutoff energy are $s=1.5$ and $\epsilon_c=15$ TeV. The total energy in the cosmic rays is $W_p= (1.07 \times 10^{49}$ erg)$\cdot d_1^2/n_1$, where $d_1$ and $n_1$ are the source distance to Earth and average density for the target material, in units of 1 kpc and 1 cm$^{-3}$, respectively. However, such a hard particle distribution is not consistent with predictions from standard cosmic ray acceleration theory. Some authors believe that, under very specific conditions in the ambient, hard gamma-ray emission can result from hadronic interactions of accelerated cosmic rays with dense, cold gas which does not emit thermal X-rays, but this scenario is not explored here. The reader is referred to the literature for a discussion about some of the corresponding models \citep[e.g.,][and references therein]{2014MNRAS.445L..70G}.

The hard GeV spectrum, on the other hand, can be naturally explained by leptonic emission. Recent analysis of LAT data from a TeV shell with a GeV spectrum similar to that of G323.7-1.0, showing a photon index of 1.4 (RCW 86), has concluded that the emission is likely leptonic \citep{2016ApJ...819...98A}. Next, using a simple one-zone leptonic model with a particle distribution described by a power law with an exponential cutoff, a set of parameters consistent with the observations is derived.

\begin{figure}[ht!]
  \includegraphics[width=10cm,height=6cm]{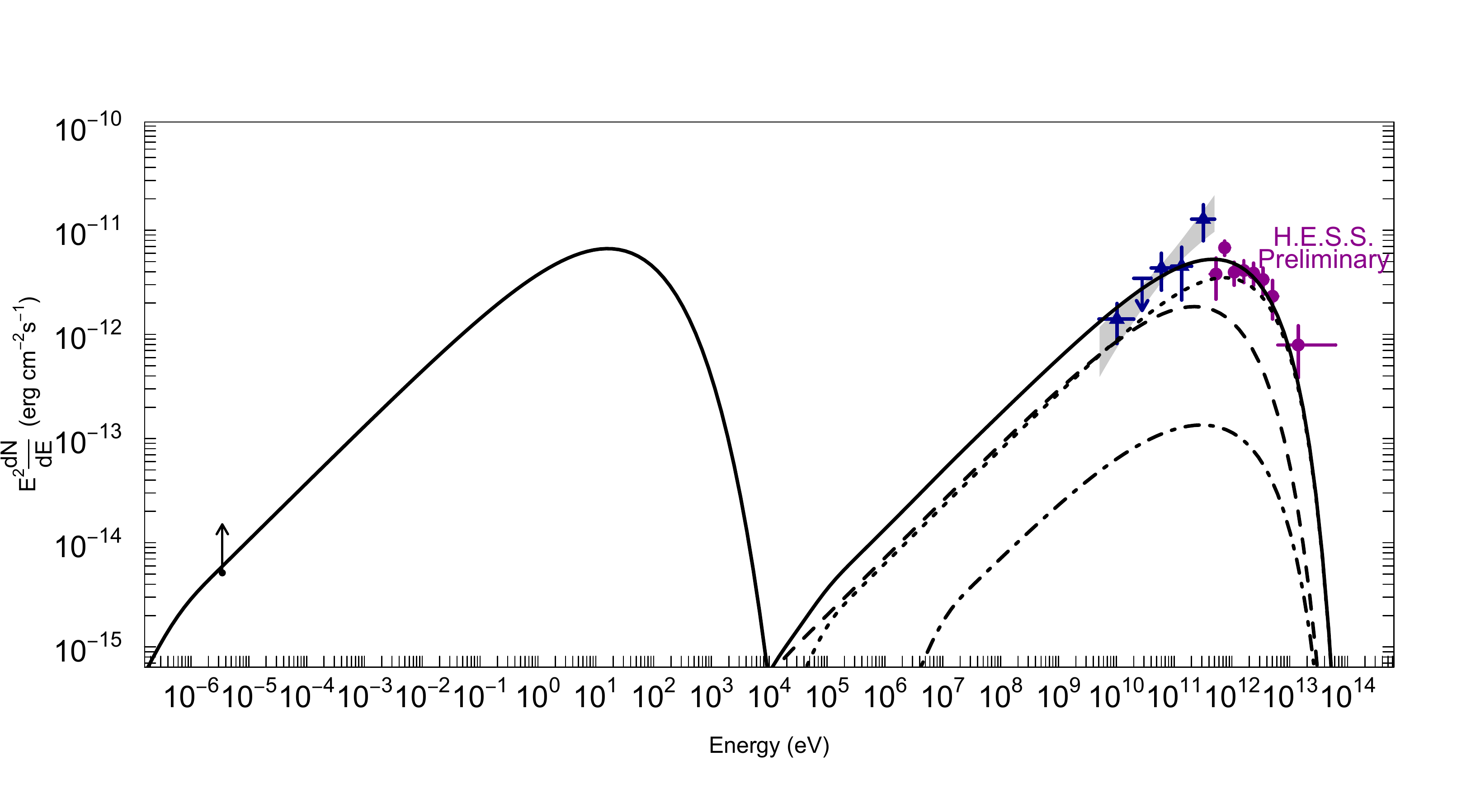}
\caption{Leptonic model for the source HESS J1534-571 showing the two characteristic ``bumps'', one at low energies corresponding to synchrotron emission from high-energy electrons and another one resulting from IC scattering of ambient photons to gamma-ray energies by the same electrons. The components of IC emission are due to interactions with the CMB (dashed line), FIR (dotted line) and NIR photons (dash-dotted line). The \emph{Fermi} LAT data from this work and the preliminary H.E.S.S. SED are shown. The lower limit on the flux from a 843 MHz radio observation taken from \cite{2014PASA...31...42G} is also shown.\label{leptonic}}
\end{figure}

Fig. \ref{leptonic} shows the SED of HESS J1534-571 with the GeV spectrum obtained here, the preliminary H.E.S.S. data points \citep{2016arXiv161200261G} and the lower limit on the radio flux at a frequency of 843 MHz from the Molonglo Galactic Plane Survey \citep{2014PASA...31...42G}. The figure also shows the leptonic emission calculated with the naima modeling package \citep{2015arXiv150903319Z} for a particle cutoff energy of 7 TeV and a spectral index of $s=1.9$. The average source magnetic field used is 6 $\mu$G. Different contributions to the gamma-ray flux from IC scattering of far infrared (FIR) photon fields from heated dust and near-infrared star light (NIR), both described by modified blackbody spectra, as well as from the Cosmic Microwave Background (CMB), are included. The contribution from NIR is negligible for typical values in the Galaxy and the temperature and density found near the Solar System are used here, 3000 K and 0.4 eV/cm$^3$\citep[e.g.,][]{2011ApJ...727...38S,2016PhRvD..94f3009V}. With the adopted model parameters, a FIR temperature and density of 20 K and 0.8 eV/cm$^3$ predict GeV-TeV fluxes that are reasonably consistent with the data. This value of the FIR density is found at a distance of 5 kpc from the Galactic center to the source using standard Galactic models \citep{2011ApJ...727...38S,2016PhRvD..94f3009V}. An estimation of the dust temperature using GALPROP \citep{2011ApJ...738...42G} for a Galactocentric distance of 5 kpc gives $\sim 30$ K, but we note that changing the dust temperature from 10 K to 50 K has little effect on the resulting IC fluxes. A Galactocentric distance of 5 kpc would place the SNR at $\sim 4-5$ kpc from the Earth, near or at the Scutum-Centaurus arm of the Milky Way \citep[e.g.,][]{2014ApJ...797...53G}. A value of 5 kpc for the distance from the source to Earth requires a total leptonic energy in the source that is very reasonable, $\sim 0.15$\% of a typical supernova kinetic explosion energy of $10^{51}$ erg, and similar to the lepton energy content in other TeV shells such as RCW 86 \citep{2016arXiv160104461H}.

Using these parameters, the model predicts a very low $2-10$ keV X-ray flux of $2.7\times 10^{-14}$ erg cm$^{-2}$ s$^{-1}$ and a very steep X-ray SED ($E^2 \frac{dN}{dE} \propto E^{-3.1}$).

The resulting nonthermal bremsstrahlung emission at gamma-ray energies is negligible for an average target density of 1 cm$^{-3}$.

Although some model parameters could vary slightly and still account for the data reasonably well, the particle distribution is likely not much harder than the one proposed here. Although a model with a harder particle distribution ($s<1.9$) better reproduces the observed GeV and TeV spectral shape, the radio flux lower limit becomes an important constraint against scenarios with harder spectra. Similarly, the particle index is likely not higher than $\sim 2.1$ because, first, that would require more energy in the electrons and, in order to keep the predicted X-ray fluxes from increasing significantly, a lower average magnetic field, which is already relatively low, but, most importantly, the shape of the predicted GeV-TeV spectrum would deviate from the observations.

HESS J1534-571 could belong to a new class of relatively evolved SNR shells that are leptonic-dominated and show hard GeV spectra. Non-thermal X-rays are expected to decrease as a SNR evolves \citep{2012ApJ...746..134N} and even though the evolution of the downstream magnetic field in a SNR is not well understood, it is expected to decrease with decreasing shock speeds \citep[e.g.,][]{2005A&A...433..229V}. Other similar objects could include the relatively faint SNR G150.3+4.5 \citep{2014A&A...567A..59G,0067-0049-222-1-5}. Follow up observations of HESS J1534-571 and modeling are necessary to confirm this scenario.

\acknowledgments

This work has received funding from the European Union's Horizon 2020 research and innovation programme under the Marie Sk\l{}odowska-Curie grant agreement No 690575. Partial financial support from Universidad de Costa Rica is also acknowledged. The comments made by the anonymous referee were very helpful to improve the quality of this paper.

\bibliographystyle{aasjournal}
\bibliography{references}

\begin{thebibliography}{}
\expandafter\ifx\csname natexlab\endcsname\relax\def\natexlab#1{#1}\fi
\providecommand{\url}[1]{\href{#1}{#1}}

\bibitem[{{Abdo} {et~al.}(2009){Abdo}, {Ackermann}, {Ajello}, {Baldini},
  {Ballet}, {Barbiellini}, {Baring}, {Bastieri}, {Baughman}, {Bechtol}, \&
  et~al.}]{2009ApJ...706L...1A}
{Abdo}, A.~A., {Ackermann}, M., {Ajello}, M., {et~al.} 2009, \apjl, 706, L1

\bibitem[{{Acero} {et~al.}(2015{\natexlab{a}}){Acero}, {Lemoine-Goumard},
  {Renaud}, {Ballet}, {Hewitt}, {Rousseau}, \& {Tanaka}}]{2015A&A...580A..74A}
{Acero}, F., {Lemoine-Goumard}, M., {Renaud}, M., {et~al.} 2015{\natexlab{a}},
  \aap, 580, A74

\bibitem[{{Acero} {et~al.}(2015{\natexlab{b}}){Acero}, {Ackermann}, {Ajello},
  {Albert}, {Atwood}, {Axelsson}, {Baldini}, {Ballet}, {Barbiellini},
  {Bastieri}, {Belfiore}, {Bellazzini}, {Bissaldi}, \&
  et~al.}]{2015ApJS..218...23A}
{Acero}, F., {Ackermann}, M., {Ajello}, M., {et~al.} 2015{\natexlab{b}}, \apjs,
  218, 23

\bibitem[{{Acero} {et~al.}(2016){Acero}, {Ackermann}, {Ajello}, {Baldini},
  {Ballet}, {Barbiellini}, {Bastieri}, {Bellazzini}, {Bissaldi}, {Blandford},
  {Bloom}, {Bonino}, {Bottacini}, {Brandt}, {Bregeon}, {Bruel}, {Buehler},
  {Buson}, {Caliandro}, {Cameron}, {Caputo}, {Caragiulo}, {Caraveo},
  {Casandjian}, {Cavazzuti}, {Cecchi}, {Chekhtman}, {Chiang}, {Chiaro},
  {Ciprini}, {Claus}, {Cohen}, {Cohen-Tanugi}, {Cominsky}, {Condon}, {Conrad},
  {Cutini}, {D'Ammando}, {de Angelis}, {de Palma}, {Desiante}, {Digel}, {Di
  Venere}, {Drell}, {Drlica-Wagner}, {Favuzzi}, {Ferrara}, {Franckowiak},
  {Fukazawa}, {Funk}, {Fusco}, {Gargano}, {Gasparrini}, {Giglietto}, {Giommi},
  {Giordano}, {Giroletti}, {Glanzman}, {Godfrey}, {Gomez-Vargas}, {Grenier},
  {Grondin}, {Guillemot}, {Guiriec}, {Gustafsson}, {Hadasch}, {Harding},
  {Hayashida}, {Hays}, {Hewitt}, {Hill}, {Horan}, {Hou}, {Iafrate}, {Jogler},
  {J{\'o}hannesson}, {Johnson}, {Kamae}, {Katagiri}, {Kataoka}, {Katsuta},
  {Kerr}, {Kn{\"o}dlseder}, {Kocevski}, {Kuss}, {Laffon}, {Lande}, {Larsson},
  {Latronico}, {Lemoine-Goumard}, {Li}, {Li}, {Longo}, {Loparco}, {Lovellette},
  {Lubrano}, {Magill}, {Maldera}, {Marelli}, {Mayer}, {Mazziotta}, {Michelson},
  {Mitthumsiri}, {Mizuno}, {Moiseev}, {Monzani}, {Moretti}, {Morselli},
  {Moskalenko}, {Murgia}, {Nemmen}, {Nuss}, {Ohsugi}, {Omodei}, {Orienti},
  {Orlando}, {Ormes}, {Paneque}, {Perkins}, {Pesce-Rollins}, {Petrosian},
  {Piron}, {Pivato}, {Porter}, {Rain{\`o}}, {Rando}, {Razzano}, {Razzaque},
  {Reimer}, {Reimer}, {Renaud}, {Reposeur}, {Rousseau}, {Saz Parkinson},
  {Schmid}, {Schulz}, {Sgr{\`o}}, {Siskind}, {Spada}, {Spandre}, {Spinelli},
  {Strong}, {Suson}, {Tajima}, {Takahashi}, {Tanaka}, {Thayer}, {Thompson},
  {Tibaldo}, {Tibolla}, {Torres}, {Tosti}, {Troja}, {Uchiyama}, {Vianello},
  {Wells}, {Wood}, {Wood}, {Yassine}, {den Hartog}, \&
  {Zimmer}}]{2016ApJS..224....8A}
---. 2016, \apjs, 224, 8

\bibitem[{{Ackermann} {et~al.}(2013){Ackermann}, {Ajello}, {Allafort},
  {Baldini}, {Ballet}, {Barbiellini}, {Baring}, {Bastieri}, {Bechtol},
  {Bellazzini}, \& et~al.}]{2013Sci...339..807A}
{Ackermann}, M., {Ajello}, M., {Allafort}, A., {et~al.} 2013, Science, 339, 807

\bibitem[{Ackermann {et~al.}(2016)Ackermann, Ajello, Atwood, Baldini, Ballet,
  Barbiellini, Bastieri, Gonzalez, Bellazzini, Bissaldi, Blandford, Bloom,
  Bonino, Bottacini, Brandt, Bregeon, Bruel, Buehler, Buson, Caliandro,
  Cameron, Caputo, Caragiulo, Caraveo, Cavazzuti, Cecchi, Charles, Chekhtman,
  Cheung, Chiang, Chiaro, Ciprini, Cohen, Cohen-Tanugi, Cominsky, Conrad,
  Cuoco, Cutini, D’Ammando, de~Angelis, de~Palma, Desiante, Mauro, Venere,
  Domínguez, Drell, Favuzzi, Fegan, Ferrara, Focke, Fortin, Franckowiak,
  Fukazawa, Funk, Furniss, Fusco, Gargano, Gasparrini, Giglietto, Giommi,
  Giordano, Giroletti, Glanzman, Godfrey, Grenier, Grondin, Guillemot, Guiriec,
  Harding, Hays, Hewitt, Hill, Horan, Iafrate, Hartmann, Jogler, Jóhannesson,
  Johnson, Kamae, Kataoka, Knödlseder, Kuss, Mura, Larsson, Latronico,
  Lemoine-Goumard, Li, Li, Longo, Loparco, Lott, Lovellette, Lubrano, Madejski,
  Maldera, Manfreda, Mayer, Mazziotta, Michelson, Mirabal, Mitthumsiri, Mizuno,
  Moiseev, Monzani, Morselli, Moskalenko, Murgia, Nuss, Ohsugi, Omodei,
  Orienti, Orlando, Ormes, Paneque, Perkins, Pesce-Rollins, Petrosian, Piron,
  Pivato, Porter, Rainò, Rando, Razzano, Razzaque, Reimer, Reimer, Reposeur,
  Romani, Sánchez-Conde, Parkinson, Schmid, Schulz, Sgrò, Siskind, Spada,
  Spandre, Spinelli, Suson, Tajima, Takahashi, Takahashi, Takahashi, Thayer,
  Thompson, Tibaldo, Torres, Tosti, Troja, Vianello, Wood, Wood, Yassine,
  Zaharijas, \& Zimmer}]{0067-0049-222-1-5}
Ackermann, M., Ajello, M., Atwood, W.~B., {et~al.} 2016, The Astrophysical
  Journal Supplement Series, 222, 5.
\newblock \url{http://stacks.iop.org/0067-0049/222/i=1/a=5}

\bibitem[{{Aharonian} \& {Atoyan}(1996)}]{1996A&A...309..917A}
{Aharonian}, F.~A., \& {Atoyan}, A.~M. 1996, \aap, 309, 917

\bibitem[{{Ajello} {et~al.}(2016){Ajello}, {Baldini}, {Barbiellini},
  {Bastieri}, {Bellazzini}, {Bissaldi}, {Bloom}, {Bonino}, {Bottacini},
  {Brandt}, {Bregeon}, {Bruel}, {Buehler}, {Caliandro}, {Cameron}, {Caragiulo},
  {Cavazzuti}, {Charles}, {Chekhtman}, {Ciprini}, {Cohen-Tanugi}, {Condon},
  {Costanza}, {Cutini}, {D'Ammando}, {de Palma}, {Desiante}, {Di Lalla}, {Di
  Mauro}, {Di Venere}, {Drell}, {Dubner}, {Dumora}, {Duvidovich}, {Favuzzi},
  {Focke}, {Fusco}, {Gargano}, {Gasparrini}, {Giacani}, {Giglietto},
  {Glanzman}, {Green}, {Grenier}, {Guiriec}, {Hays}, {Hewitt}, {Hill}, {Horan},
  {Jogler}, {J{\'o}hannesson}, {Jung-Richardt}, {Kensei}, {Kuss}, {Larsson},
  {Latronico}, {Lemoine-Goumard}, {Li}, {Li}, {Longo}, {Loparco}, {Lovellette},
  {Lubrano}, {Magill}, {Maldera}, {Manfreda}, {Mayer}, {Mazziotta}, {McEnery},
  {Michelson}, {Mitthumsiri}, {Mizuno}, {Monzani}, {Morselli}, {Moskalenko},
  {Negro}, {Nuss}, {Orienti}, {Orlando}, {Ormes}, {Paneque}, {Perkins},
  {Pesce-Rollins}, {Piron}, {Pivato}, {Porter}, {Rain{\`o}}, {Rando},
  {Razzano}, {Reimer}, {Reimer}, {Reposeur}, {Schmid}, {Schulz}, {Sgr{\`o}},
  {Simone}, {Siskind}, {Spada}, {Spandre}, {Spinelli}, {Thayer}, {Tibaldo},
  {Torres}, {Tosti}, {Troja}, {Uchiyama}, {Vianello}, {Vink}, {Wood}, \&
  {Yassine}}]{2016ApJ...819...98A}
{Ajello}, M., {Baldini}, L., {Barbiellini}, G., {et~al.} 2016, \apj, 819, 98

\bibitem[{{Araya} \& {Cui}(2010)}]{2010ApJ...720...20A}
{Araya}, M., \& {Cui}, W. 2010, \apj, 720, 20

\bibitem[{{Araya} \& {Frutos}(2012)}]{2012MNRAS.425.2810A}
{Araya}, M., \& {Frutos}, F. 2012, \mnras, 425, 2810

\bibitem[{{Atwood} {et~al.}(2009){Atwood}, {Abdo}, {Ackermann}, {Althouse},
  {Anderson}, {Axelsson}, {Baldini}, {Ballet}, {Band}, {Barbiellini}, \&
  et~al.}]{2009ApJ...697.1071A}
{Atwood}, W.~B., {Abdo}, A.~A., {Ackermann}, M., {et~al.} 2009, \apj, 697, 1071

\bibitem[{{Bell}(1978)}]{1978MNRAS.182..147B}
{Bell}, A.~R. 1978, \mnras, 182, 147

\bibitem[{{Bell}(2004)}]{2004MNRAS.353..550B}
---. 2004, \mnras, 353, 550

\bibitem[{{Capasso} {et~al.}(2016){Capasso}, {Condon}, {Coffaro}, {Cui},
  {Gottschall}, {Klochkov}, {Marandon}, {Maxted}, {P{\"u}hlhofer}, {Rowell}, \&
  {for the H.~E.~S.~S.~Collaboration}}]{2016arXiv161200258C}
{Capasso}, M., {Condon}, B., {Coffaro}, M., {et~al.} 2016, ArXiv e-prints,
  arXiv:1612.00258

\bibitem[{{Fukui} {et~al.}(2012){Fukui}, {Sano}, {Sato}, {Torii}, {Horachi},
  {Hayakawa}, {McClure-Griffiths}, {Rowell}, {Inoue}, {Inutsuka}, {Kawamura},
  {Yamamoto}, {Okuda}, {Mizuno}, {Onishi}, {Mizuno}, \&
  {Ogawa}}]{2012ApJ...746...82F}
{Fukui}, Y., {Sano}, H., {Sato}, J., {et~al.} 2012, \apj, 746, 82

\bibitem[{{Gabici} \& {Aharonian}(2014)}]{2014MNRAS.445L..70G}
{Gabici}, S., \& {Aharonian}, F.~A. 2014, \mnras, 445, L70

\bibitem[{{Gao} \& {Han}(2014)}]{2014A&A...567A..59G}
{Gao}, X.~Y., \& {Han}, J.~L. 2014, \aap, 567, A59

\bibitem[{{Goodman} {et~al.}(2014){Goodman}, {Alves}, {Beaumont}, {Benjamin},
  {Borkin}, {Burkert}, {Dame}, {Jackson}, {Kauffmann}, {Robitaille}, \&
  {Smith}}]{2014ApJ...797...53G}
{Goodman}, A.~A., {Alves}, J., {Beaumont}, C.~N., {et~al.} 2014, \apj, 797, 53

\bibitem[{{Gottschall} {et~al.}(2016){Gottschall}, {Capasso}, {Deil},
  {Djannati-Atai}, {Donath}, {Eger}, {Marandon}, {Maxted}, {P{\"u}hlhofer},
  {Renaud}, {Sasaki}, {Terrier}, {Vink}, \& {for the
  H.~E.~S.~S.~Collaboration}}]{2016arXiv161200261G}
{Gottschall}, D., {Capasso}, M., {Deil}, C., {et~al.} 2016, ArXiv e-prints,
  arXiv:1612.00261

\bibitem[{{Green} {et~al.}(2014){Green}, {Reeves}, \&
  {Murphy}}]{2014PASA...31...42G}
{Green}, A.~J., {Reeves}, S.~N., \& {Murphy}, T. 2014, \pasa, 31, e042

\bibitem[{{Grondin} {et~al.}(2011){Grondin}, {Funk}, {Lemoine-Goumard}, {Van
  Etten}, {Hinton}, {Camilo}, {Cognard}, {Espinoza}, {Freire}, {Grove},
  {Guillemot}, {Johnston}, {Kramer}, {Lande}, {Michelson}, {Possenti},
  {Romani}, {Skilton}, {Theureau}, \& {Weltevrede}}]{2011ApJ...738...42G}
{Grondin}, M.-H., {Funk}, S., {Lemoine-Goumard}, M., {et~al.} 2011, \apj, 738,
  42

\bibitem[{{H.~E.~S.~S.~Collaboration}
  {et~al.}(2016){H.~E.~S.~S.~Collaboration}, {Abramowski}, {Aharonian}, {Ait
  Benkhali}, {Akhperjanian}, {Ang{\"u}ner}, {Backes}, {Balzer}, {Becherini},
  {Becker Tjus}, \& et~al.}]{2016arXiv160104461H}
{H.~E.~S.~S.~Collaboration}, {Abramowski}, A., {Aharonian}, F., {et~al.} 2016,
  ArXiv e-prints, arXiv:1601.04461

\bibitem[{{Jogler} \& {Funk}(2016)}]{2016ApJ...816..100J}
{Jogler}, T., \& {Funk}, S. 2016, \apj, 816, 100

\bibitem[{{Kafexhiu} {et~al.}(2014){Kafexhiu}, {Aharonian}, {Taylor}, \&
  {Vila}}]{2014PhRvD..90l3014K}
{Kafexhiu}, E., {Aharonian}, F., {Taylor}, A.~M., \& {Vila}, G.~S. 2014, \prd,
  90, 123014

\bibitem[{{Krymskii}(1977)}]{1977DoSSR.234.1306K}
{Krymskii}, G.~F. 1977, Akademiia Nauk SSSR Doklady, 234, 1306

\bibitem[{{Mattox} {et~al.}(1996){Mattox}, {Bertsch}, {Chiang}, {Dingus},
  {Digel}, {Esposito}, {Fierro}, {Hartman}, {Hunter}, {Kanbach}, \&
  et~al.}]{1996ApJ...461..396M}
{Mattox}, J.~R., {Bertsch}, D.~L., {Chiang}, J., {et~al.} 1996, \apj, 461, 396

\bibitem[{{Nakamura} {et~al.}(2012){Nakamura}, {Bamba}, {Dotani}, {Ishida},
  {Yamazaki}, \& {Kohri}}]{2012ApJ...746..134N}
{Nakamura}, R., {Bamba}, A., {Dotani}, T., {et~al.} 2012, \apj, 746, 134

\bibitem[{{P{\"u}hlhofer} {et~al.}(2015){P{\"u}hlhofer}, {Brun}, {Capasso},
  {Chaves}, {Deil}, {Djannati-Ata{\"i}}, {Donath}, {Eger}, {Gottschall},
  {Laffon}, {Marandon}, {Oakes}, {Renaud}, {Sasaki}, {Terrier}, {Vink}, {for
  the H.~E.~S.~S.~collaboration}, \& {Bamba}}]{2015arXiv150903872P}
{P{\"u}hlhofer}, G., {Brun}, F., {Capasso}, M., {et~al.} 2015, ArXiv e-prints,
  arXiv:1509.03872

\bibitem[{{Sakai} {et~al.}(2011){Sakai}, {Yajima}, \&
  {Matsumoto}}]{2011PASJ...63S.879S}
{Sakai}, M., {Yajima}, Y., \& {Matsumoto}, H. 2011, \pasj, 63, S879

\bibitem[{{Sano} {et~al.}(2015){Sano}, {Fukuda}, {Yoshiike}, {Sato}, {Horachi},
  {Kuwahara}, {Torii}, {Hayakawa}, {Tanaka}, {Matsumoto}, {Inoue}, {Yamazaki},
  {Inutsuka}, {Kawamura}, {Yamamoto}, {Okuda}, {Tachihara}, {Mizuno}, {Onishi},
  {Mizuno}, {Acero}, \& {Fukui}}]{2015ApJ...799..175S}
{Sano}, H., {Fukuda}, T., {Yoshiike}, S., {et~al.} 2015, \apj, 799, 175

\bibitem[{{Shibata} {et~al.}(2011){Shibata}, {Ishikawa}, \&
  {Sekiguchi}}]{2011ApJ...727...38S}
{Shibata}, T., {Ishikawa}, T., \& {Sekiguchi}, S. 2011, \apj, 727, 38

\bibitem[{{Takeda} {et~al.}(2016){Takeda}, {Bamba}, {Terada}, {Tashiro},
  {Katsuda}, {Yamazaki}, {Ohira}, \& {Iwakiri}}]{2016PASJ...68S..10T}
{Takeda}, S., {Bamba}, A., {Terada}, Y., {et~al.} 2016, \pasj, 68, S10

\bibitem[{{The Fermi-LAT Collaboration}(2017)}]{2017arXiv170200664T}
{The Fermi-LAT Collaboration}. 2017, ArXiv e-prints, arXiv:1702.00664

\bibitem[{{The Fermi LAT Collaboration} {et~al.}(2017){The Fermi LAT
  Collaboration}, {Ackermann}, {Ajello}, {Baldini}, {Ballet}, {Barbiellini},
  {Bastieri}, {Bellazzini}, {Bissaldi}, {Bloom}, {Bonino}, {Bottacini},
  {Brandt}, {Bregeon}, {Bruel}, {Buehler}, {Cameron}, {Caragiulo}, {Caraveo},
  {Castro}, {Cavazzuti}, {Cecchi}, {Charles}, {Chekhtman}, {Cheung}, {Chiaro},
  {Ciprini}, {Cohen}, {Costantin}, {Costanza}, {Cutini}, {D'Ammando}, {de
  Palma}, {Desiante}, {Digel}, {Di Lalla}, {Di Mauro}, {Di Venere}, {Favuzzi},
  {Fegan}, {Ferrara}, {Franckowiak}, {Fukazawa}, {Funk}, {Fusco}, {Gargano},
  {Gasparrini}, {Giglietto}, {Giordano}, {Giroletti}, {Green}, {Grenier},
  {Grondin}, {Guillemot}, {Guiriec}, {Harding}, {Hays}, {Hewitt}, {Horan},
  {Hou}, {J{\'o}hannesson}, {Kamae}, {Kuss}, {La Mura}, {Larsson},
  {Lemoine-Goumard}, {Li}, {Longo}, {Loparco}, {Lubrano}, {Magill}, {Maldera},
  {Malyshev}, {Manfreda}, {Mazziotta}, {Michelson}, {Mitthumsiri}, {Mizuno},
  {Monzani}, {Morselli}, {Moskalenko}, {Negro}, {Nuss}, {Ohsugi}, {Omodei},
  {Orienti}, {Orlando}, {Ormes}, {Paliya}, {Paneque}, {Perkins}, {Persic},
  {Pesce-Rollins}, {Petrosian}, {Piron}, {Porter}, {Principe}, {Raino},
  {Rando}, {Razzano}, {Razzaque}, {Reimer}, {Reimer}, {Reposeur}, {Sgro},
  {Simone}, {Siskind}, {Spada}, {Spandre}, {Spinelli}, {Suson}, {Tak},
  {Thayer}, {Thompson}, {Torres}, {Tosti}, {Troja}, {Vianello}, {Wood}, \&
  {Wood}}]{2017arXiv170200476T}
{The Fermi LAT Collaboration}, {Ackermann}, M., {Ajello}, M., {et~al.} 2017,
  ArXiv e-prints, arXiv:1702.00476

\bibitem[{{Vernetto} \& {Lipari}(2016)}]{2016PhRvD..94f3009V}
{Vernetto}, S., \& {Lipari}, P. 2016, \prd, 94, 063009

\bibitem[{{V{\"o}lk} {et~al.}(2005){V{\"o}lk}, {Berezhko}, \&
  {Ksenofontov}}]{2005A&A...433..229V}
{V{\"o}lk}, H.~J., {Berezhko}, E.~G., \& {Ksenofontov}, L.~T. 2005, \aap, 433,
  229

\bibitem[{{Zabalza}(2015)}]{2015arXiv150903319Z}
{Zabalza}, V. 2015, Proceedings of the 34th International Cosmic Ray Conference
  (ICRC2015), The Hague, The Netherlands, arXiv:1509.03319

\end{thebibliography}

\end{document}